\documentclass[a4paper]{jpconf}
\usepackage{graphicx}
\begin{document}
\title{Cooperative spontaneous decay of local excitation in a dense and disordered ensemble of point-like impurity atoms near a charged conductive surface}

\author{A S Kuraptsev}

\address{Peter the Great St. Petersburg Polytechnic University, 195251, St. Petersburg, Russia}

\ead{aleksej-kurapcev@yandex.ru}

\begin{abstract}
On the basis of the general quantum microscopic theory we study the process of spontaneous decay of an excited atom in a dense and disordered ensemble of point-like impurity atoms embedded into transparent dielectric and located near a charged perfectly conducting surface. We have analyzed the simultaneous influence of the modified spatial structure of field modes near the conductive surface and the electric field on the character of interatomic dipole-dipole interaction. This leads to the modification of the transition spectrum of an excited atom inside an ensemble and the spontaneous decay dynamics. We have shown that the electric field changes the cooperative Lamb shift, as well as the character of sub- and superradiant decay.
\end{abstract}

\section{Introduction}
Since the pioneering work of Purcell \cite{1}, the interaction of light
with atoms localized inside a cavity or waveguide, as well as
near its surface, has attracted considerable attention. Now it is
well understood that a cavity modifies the spatial structure of the
modes of the electromagnetic field. This leads to the modification
of the radiative properties of atoms, and in particular to
the enhancement and inhibition of the spontaneous decay rate
\cite{2} -- \cite{5}. This proposes an exciting tool for the preparation of media with given optical properties.
Modification in the structure of field modes changes not only single-particle characteristics but also the character of photon exchange between different atoms. In
its turn this leads to an alteration of the dipole-dipole interatomic
interaction \cite{15} -- \cite{16}, as well as associated cooperative
effects \cite{17} -- \cite{Ruo1}.

In fact, not only cavity or waveguide can modify the spatial structure of the modes of the electromagnetic
field. Single metallic surface also has this property. For this
reason, the characteristics of the ensemble of atoms or quantum dots located
near the conductive surface differ from ones in the case of
the same ensemble in free space \cite{31}. If the
metallic surface is charged, an electrostatic field causes Stark shifts of the atomic energy levels, which leads to additional
modification of the interatomic dipole-dipole interaction \cite{KS_Laser_Phys_2018} -- \cite{KS_JETP_2018}.

The main goal of this work is to describe theoretically the influence of the dipole-dipole interaction in a dense ensemble of pointlike impurity
atoms embedded in a solid dielectric and placed near a perfectly conductive charged plate on the spontaneous decay dynamics of the spatially localized atomic excitation prepared inside this ensemble. We simultaneously analyze two factors affecting the character of spontaneous decay in the polyatomic system with strong interatomic correlations: the peculiarities of the spatial structure of field modes near the conductive surface as well as Stark splitting of energy levels induced by an electrostatic field.

\section{Basic assumptions and approach}
We consider an ensemble, which consists of $N\gg1$ motionless
impurity atoms embedded in a transparent dielectric and
placed near a charged perfectly conducting plate. The longitudinal dimensions of the plate are much larger than the wavelength of light corresponding to exact resonance with atomic transition, $\lambda_{0}$, and the sizes of a sample. It will allow us to suppose that these dimensions are infinite.

We also assume that the temperature of the dielectric is low, so that the effect of electron-phonon interaction can be neglected \cite{Gl1} -- \cite{Gl2}. Nevertheless, the internal fields of a dielectric cause shifts of spectral lines of impurity atoms even in the case
of a transparent dielectric. Therefore, we account, that the transition frequency of
impurity atoms located in a dielectric $\omega_{e_{a}}$ differs from one of a free atom $\omega_{0}$. Considering the shift caused by the influence of the dielectric matrix, $\omega_{e_{a}}=\omega_{0}+\Delta_{e_{a}}$, where $\Delta_{e_{a}}$ is the shift of transition frequency of the atom $a$ ($a = 1,...,N$) which depends on
its spatial location owing to the inhomogeneity of internal fields in
dielectric.

All these assumptions allow us to consider dynamics of the
model system which consists of the set of motionless pointlike
scatterers and electromagnetic field near ideal mirror. We take
into account all the modes of the field including the modes which are
initially in the vacuum state. During the evolution these modes
can be populated as a result of atomic decay, i.e., as a result
of interaction of the excited atom with the field in the vacuum
state. The photon created in such atomic transitions can be
absorbed by another atom in the ensemble. This atom emits
a secondary photon and so on. Thus, in our theory we deal
with the closed quantum system, which can be described by
the wave function.

In this work we use the quantum microscopic approach, which is based on the solution of the non-steady-state Schrodinger equation for the wave function of the combined system, which consists of all the impurity atoms and the electromagnetic field, including vacuum reservoir. This basic approach was
described first in \cite{Heitler} and developed afterwards in \cite{SKH_JETP_2011} for a
description of collective effects in dense and cold nondegenerate
atomic gases. It was successfully used
for a description of optical properties of dense atomic ensembles
\cite{KS_PRA_2011} -- \cite{KS_PRA_2014} and for the investigation of light scattering from these
ensembles \cite{S_J_Mod_Opt_2010} -- \cite{KS_PRA_2017}. Furthermore, we generalized the quantum microscopic approach on the case of atomic systems located in a Fabry-Perot cavity \cite{KS_PRA_2016}, \cite{KS_JETP_2016}. Mathematical formalism developed for a Fabry-Perot cavity also allowed us to analyze the dipole-dipole interaction between two motionless point atoms near a single perfectly conducting mirror \cite{KS_SPIE_2018}, \cite{KS_Laser_Phys_2018}  -- \cite{KS_JETP_2018}. Therefore, in this section we do not reproduce the general theory in detail. The reader is referred to the mentioned
papers for the theoretical developments and justifications. In the following paragraphs, we just provide a brief overview of the approach and explain, how the electric field can be taken into account.

Full Hamiltonian $\widehat{H}$ of the joint system can be written as follows:
\begin{equation}\label{1}
\widehat{H}=\widehat{H}_{0}+\widehat{V},
\end{equation}
\begin{equation}\label{1a}
\widehat{H}_{0}=\widehat{H}_{f}+\sum_{a}\widehat{H}_{a}+\sum_{a}\widehat{H}_{a\mathcal{E}}.
\end{equation}
Here $\widehat{H}_{a}$ is the Hamiltonian of the atom \emph{a} noninteracting with the field, $\widehat{H}_{f}$ is the Hamiltonian of the free field in a Fabry-Perot cavity, $\widehat{H}_{a\mathcal{E}}$ is the operator of atomic interaction with constant electric field $\bf{\mathcal{E}}$ created by charged metallic surface and $V$ is the operator of atomic interaction with oscillating field, including vacuum reservoir. The operator $V$ can be presented in the dipole approximation.

It is known that even in the case of a single atom, there are no stationary atomic states in an electric field. The electric field makes all the states decaying. This effect is known as cold electron emission. However, the process of cold emission is very slow, the typical decay rate is much less than radiative decay rate. It allows us to consider the operator $\widehat{H}_{a\mathcal{E}}$ in the framework of the perturbation theory. In such way, we can describe the joint atom-field system using the model Hamiltonian
\begin{equation}\label{1b}
\widehat{H}_{0M}=\widehat{H}_{f}+\sum_{a}\widehat{H}_{a}
\end{equation}
in the Eq. (\ref{1}) instead of $\widehat{H}_{0}$ and taking into account Stark shifts of the atomic energy levels caused by the electric field.

Let us decompose the wave function of the joint system in a set of
eigenfunctions of the operator $\widehat{H}_{0M}$. Using this decomposition, we can convert the Schrodinger equation to the system of linear differential equations for
the amplitudes of quantum states. This set of equations is infinite because of the infinity number of the field states.

The key simplification in the approach is that we
restrict the total number of quantum states, which are taken into account. We suppose that the initial excitation is weak, so that all nonlinear effects are negligible. With the accuracy up to the second order
of the fine structure constant, we can consider only the following states with no more than one photon (see \cite{Stephen_1964}):

(1) Onefold atomic excited states,
$\psi_{e_{a}}=|g,...,g,e,g,...,g\rangle\otimes|vac\rangle$, $E_{e_{a}}=\hbar\omega_{e_{a}}$.

(2) Resonant single-photon states,
$\psi_{g}=|g,...,g\rangle\otimes|\textbf{k},\alpha\rangle$, $E_{g}=\hbar\omega_{k}$.

(3) Nonresonant states. There are two excited atoms and one photon,

$\psi_{e_{a}e_{b}}=|g,...,g,e,g,...,g,e,g,...,g\rangle\otimes|\textbf{k},\alpha\rangle,
E_{e_{a}e_{b}}=\hbar(\omega_{e_{a}}+\omega_{e_{b}})+\hbar\omega_{k}$.

After the performed restriction of the total number of quantum states,
the system of equations remains infinite. However, we can, formally
solve it without any additional approximations. For this we
express the amplitudes of quantum states with one photon
via the amplitudes of atomic excitation. Then we put it in the
equations for the amplitudes of atomic excitation. Thus, we obtain
a closed finite system of equations for onefold excited states of
atomic subsystem $b_{e}$. For Fourier components $b_{e}(\omega)$ we have
(for details see \cite{SKH_JETP_2011}, \cite{KS_JETP_2016})
\begin{equation}\label{2}
\sum_{e'}\bigl[(\omega-\omega_{e'_{a}})\delta_{ee'}-\Sigma_{ee'}(\omega)\bigl]b_{e'}(\omega)=i\delta_{e s}.
\end{equation}

When deriving this expression, we assumed that, initially only one atom was excited to
a state denoted by index $s$, while all the other atoms were
in the ground state. The electromagnetic field at $t=0$
is in the vacuum state. The matrix $\Sigma_{ee'}(\omega)$ contains information about both single-atom spontaneous decay rate and
the dipole-dipole interaction between different atoms. It plays a key role in
the microscopic theory. The explicit expressions for the elements of this matrix
corresponding to a Fabry-Perot cavity were derived in Refs. \cite{KS_PRA_2016}, \cite{KS_JETP_2016}.

The dimension of the system (\ref{2}) is determined by the total number of
atoms $N$, as well as the structure of atomic energy levels. In this work
we consider the impurity atoms, which have the ground state characterizing by the total angular momentum $J=0$.
The excited state is characterized by $J=1$. It
consists of three Zeeman sublevels $e=|J, m\rangle$, which differ by the projection of the
angular momentum on the quantization axis $z$: $m =-1,0,1$. Thus, the total number of the onefold atomic excited states is equal to $3N$.

Owing to an electrostatic field $\bf{\mathcal{E}}$ of a charged plate, resonant frequencies of atomic transitions $\omega_{e_{a}}$ differ from those of an isolated atom in the free space $\omega_{0}$. Let us consider, that the quantization axis $z$ is perpendicular to the mirrors of a Fabry-Perot cavity. Taking into account Stark shift, we denote $\omega_{m=\pm1}$ the resonant frequency of transition $J=0$ $\leftrightarrow$ $J=1, m=\pm1$; $\omega_{m=0}$ -- $J=0$ $\leftrightarrow$ $J=1, m=0$ (the Lamb shift is considered to be included in the transition frequency). The influence of an electrostatic field on the character of photon exchange is significant when Stark splitting $\Delta=\omega_{m=0}-\omega_{m=\pm1}$ is comparable with natural linewidth $\gamma_{0}$. Note that this criterion is satisfied for sufficiently large values of $\mathcal{E}$.

Numerical solution of the system (\ref{2}) allows one to obtain the Fourier amplitudes of atomic excited states $b_{e}(\omega)$. Knowing $b_{e}(\omega)$, we obtain the amplitudes of all the quantum states under consideration (see \cite{SKH_JETP_2011}) and, subsequently, the wave function
of the joint atom-field system.

To analyze the dynamics of atomic ensemble located near a single mirror on the basis of mathematical formalism developed for a cavity, we should go to the limit of infinite separation between the mirrors and consider atoms near the first mirror. In this case the influence of the second mirror on the dynamics of atomic ensemble can be neglected.

In the next
section, we use the approach described here to investigate the simultaneous
influence of the peculiarities of the spatial structure of field modes near the conductive surface as well as Stark splitting of energy levels induced by an electrostatic field on the character of many-body cooperative effects. We will study the transition spectrum of the excited
atom, which is located in the ensemble of unexcited atoms, as well as the spontaneous decay dynamics of this atom.

\section{Results and discussion}
Before the analysis of many-body cooperative effects we should make some notes about single-atom properties in the presence of a charged conducting surface. When an atom is located close to the surface, the spectrum of atomic transition represents a Lorentz
profile whose width $\gamma$ depends on the distance $z_{1}$
between the atom and the surface. Accordingly, the spontaneous decay dynamics of an excited
atom is described by a single-exponential law, $P_{s}(t)=\exp(-\gamma t)$. The function $\gamma(z_{1})$ depends on Zeeman sublevel, which is initially populated. For instance, if $z_{1}=1$ (hereafter, we take $k_{0}^{-1}=\lambda_{0}/2\pi$ as the unit of length), single-atom spontaneous decay rate is $0.65 \gamma_{0}$ for Zeeman sublevels $m=\pm 1$ and $1.65 \gamma_{0}$ for $m=0$ \cite{KS_JETP_2018}. The electrostatic field does not affect the rate
of single-atom spontaneous decay, because it can only
lead to the shift of the atomic levels and does not
change their width.

Now, let us consider atomic ensemble located near a charged conductive surface. In this work we suppose that initial excitation
of the ensemble is spatially localized. This can be prepared
using the method of the two-photon resonance. In the framework of
this method the medium is illuminated by two
orthogonally propagated narrow light beams. Both beams are far-off-resonant, but their combined effect
on the atoms located in the crossing region cause two-photon excitation (of course, the
conditions of two-photon resonance must be satisfied). This method allows one to obtain a small
cluster of excited atoms in the bulk region of a dielectric. For
simplicity we will suppose that only one atom in an ensemble was excited initially.

In the case when an atom is located near the conducting surface, the natural linewidth of atomic transition depends on $z$ position
of the atom. So all the results obtained for an ensemble must
depend on $z$ position of the initially excited atom. Further we will consider
$z_{exc}=1$ (reference point $z=0$ corresponds to the position of the surface).

Note that any physical
characteristic of the ensemble with given configuration depends on the positions of all the atoms. In
this work we study disordered atomic ensembles
with uniform (on average) spatial distribution of atomic density, because it is the most typical situation for real experiments. Therefore, we
average all the calculated observables over random spatial configurations of the
ensemble by a Monte Carlo method. To take into account the inhomogeneous broadening, we also perform Monte Carlo averaging over random frequency shifts of atomic transitions caused
by the inhomogeneity of the internal fields of a dielectric.

\begin{figure}\center
	\includegraphics[width=6cm]{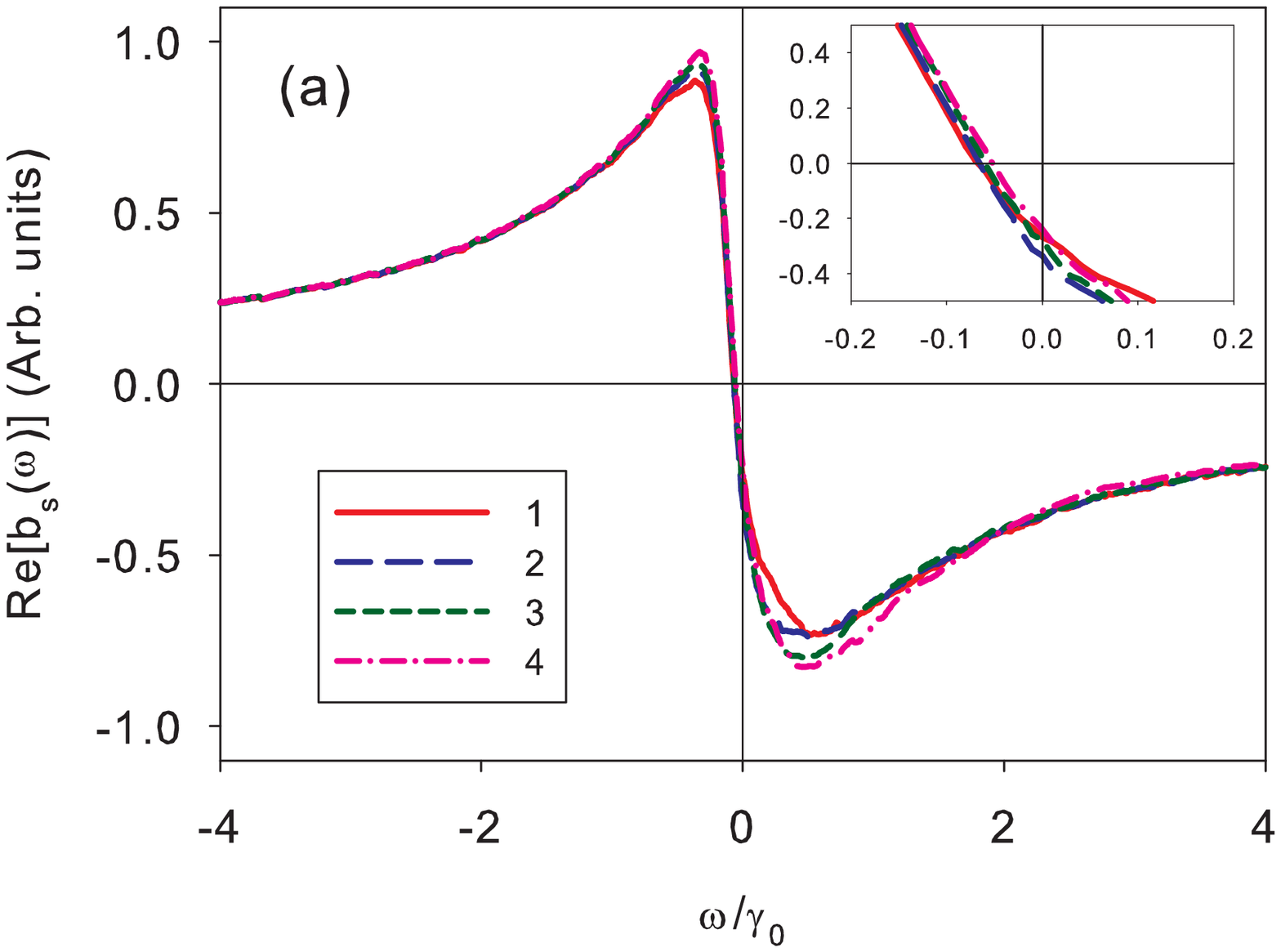}
	\includegraphics[width=6cm]{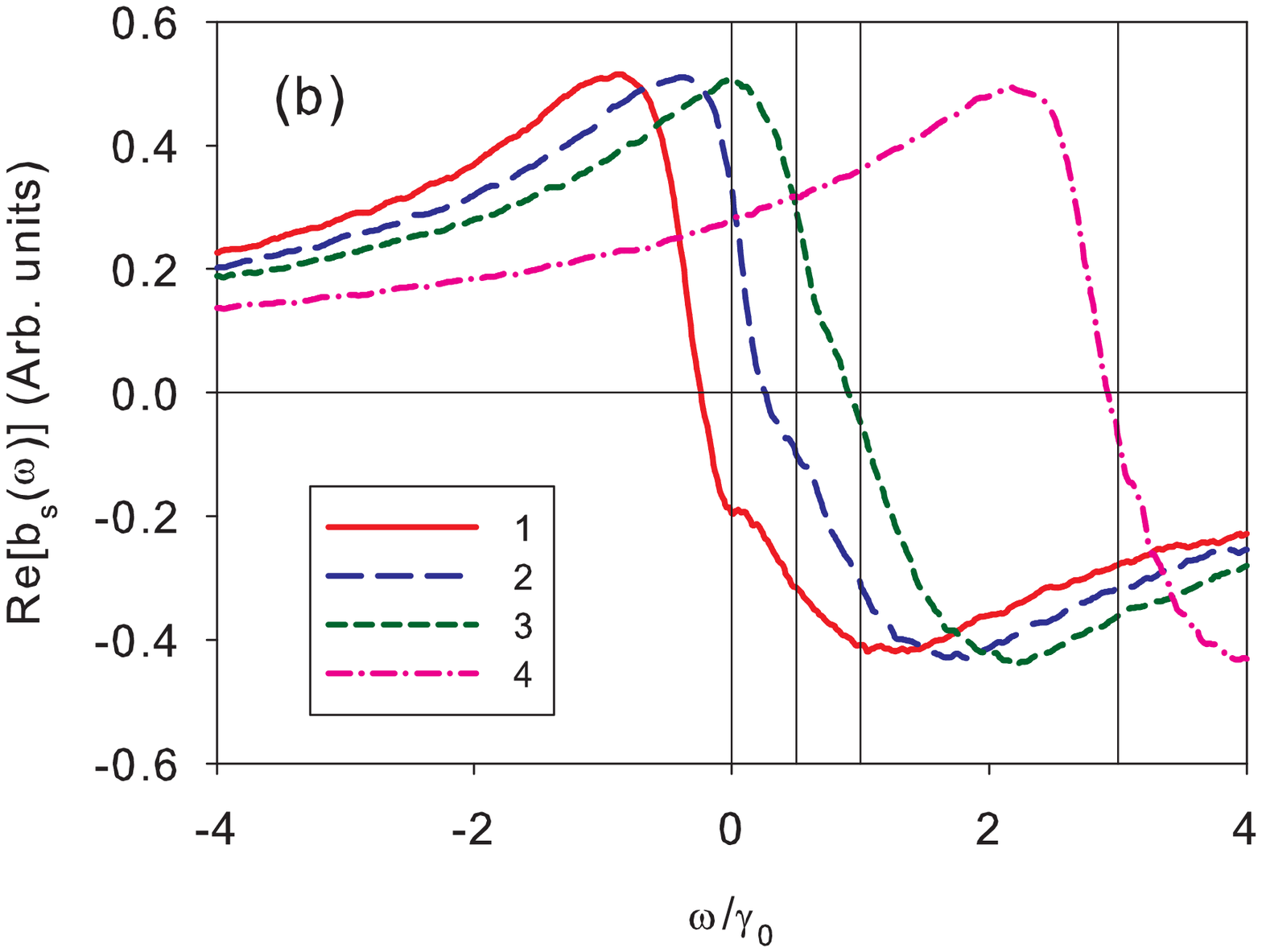}\\
	\caption{\label{fig:one}
			Transition spectrum of an atom near a charged conducting surface. $z_{exc}=1$, $n=0.05$, $\delta=0$. (a) $m=\pm1$; (b) $m=0$. 1, $\Delta=0$; 2, $\Delta=0.5\gamma_{0}$; 3, $\Delta=\gamma_{0}$; 4, $\Delta=3\gamma_{0}$.}\label{f1}
\end{figure}

Figure \ref{fig:one} shows the spectrum of the real part of the quantum amplitude of the excited state related to the initially excited atom. The calculations were carried out for the inhomogeneous broadening $\delta=0$. In this case all the atoms are mutually resonant, so
the dipole-dipole interaction manifests itself most clearly. The atomic density was chosen $n=0.05$. It is sufficiently large value, so that the
dipole-dipole interaction plays a significant role for the atomic
ensembles of such density in free space, without a surface \cite{KS_PRA_2014}. The frequency $\omega$ is counted from the resonant frequency of the transition $J=0$ $\leftrightarrow$ $J=1, m=\pm1$ of a single atom, $\omega_{m=\pm1}$. So the resonant frequency of the transition $J=0$ $\leftrightarrow$ $J=1, m=0$ of a single atom is shifted to the blue spectral area by the amount of Stark splitting $\Delta$. Therefore, in the figure \ref{fig:one}(b) we added reference vertical lines, which indicate all the considered values of Stark splitting: $\Delta=0$, $\Delta=0.5\gamma_{0}$, $\Delta=\gamma_{0}$ and $\Delta=3\gamma_{0}$.

In the figure \ref{fig:one} we see the cooperative Lamb shift (CLS) and some distortion of the shape of the spectrum. For $m=0$ the cooperative Lamb shift is larger than for $m=\pm1$, and its dependence on the electric field manifests itself more clearly. It can be also noticed, that the distortion of the spectrum shape for $m=0$ is stronger than for $m=\pm1$. In our opinion, the main reason of these peculiarities is that single-atom spontaneous decay rate for $m=0$ is 2.5 times bigger than for $m=\pm1$. So the shifts of the frequencies of collective states related to the transition $J=0$ $\leftrightarrow$ $J=1, m=0$ are larger than ones related to the transition $J=0$ $\leftrightarrow$ $J=1, m=\pm1$.

Besides CLS and distortion of the shape of the spectrum, we observe a significant broadening caused by the dipole-dipole interactions. There is also some dependence of the broadening on the electric field, but this dependence is small. It is connected with the fact, that the electric field strongly affects the energy of the collective states and weakly affects their width, as it was shown in Ref. \cite{KS_JETP_2018} for a diatomic quasimolecule.

In the general case, the specific type of transition spectrum of an excited atom in an ensemble with a given density
depends on the dimensions of this ensemble. We have analyzed how the transition spectrum changes with the sizes and found that for small ensembles,
when the mean free path of the photon is less or comparable with
linear dimensions of atomic ensemble, these changes are very significant. As linear dimensions increase, the changes in the transition spectrum become
weaker and weaker. Size dependence has a clear tendency
to saturation. The results presented in the figure \ref{fig:one} correspond to sufficiently large sample, when size dependence can be neglected. So it can be valid for a description of the transition spectrum of an excited atom
in atomic ensemble with macroscopic dimensions.

\begin{figure}\center
	\includegraphics[width=6cm]{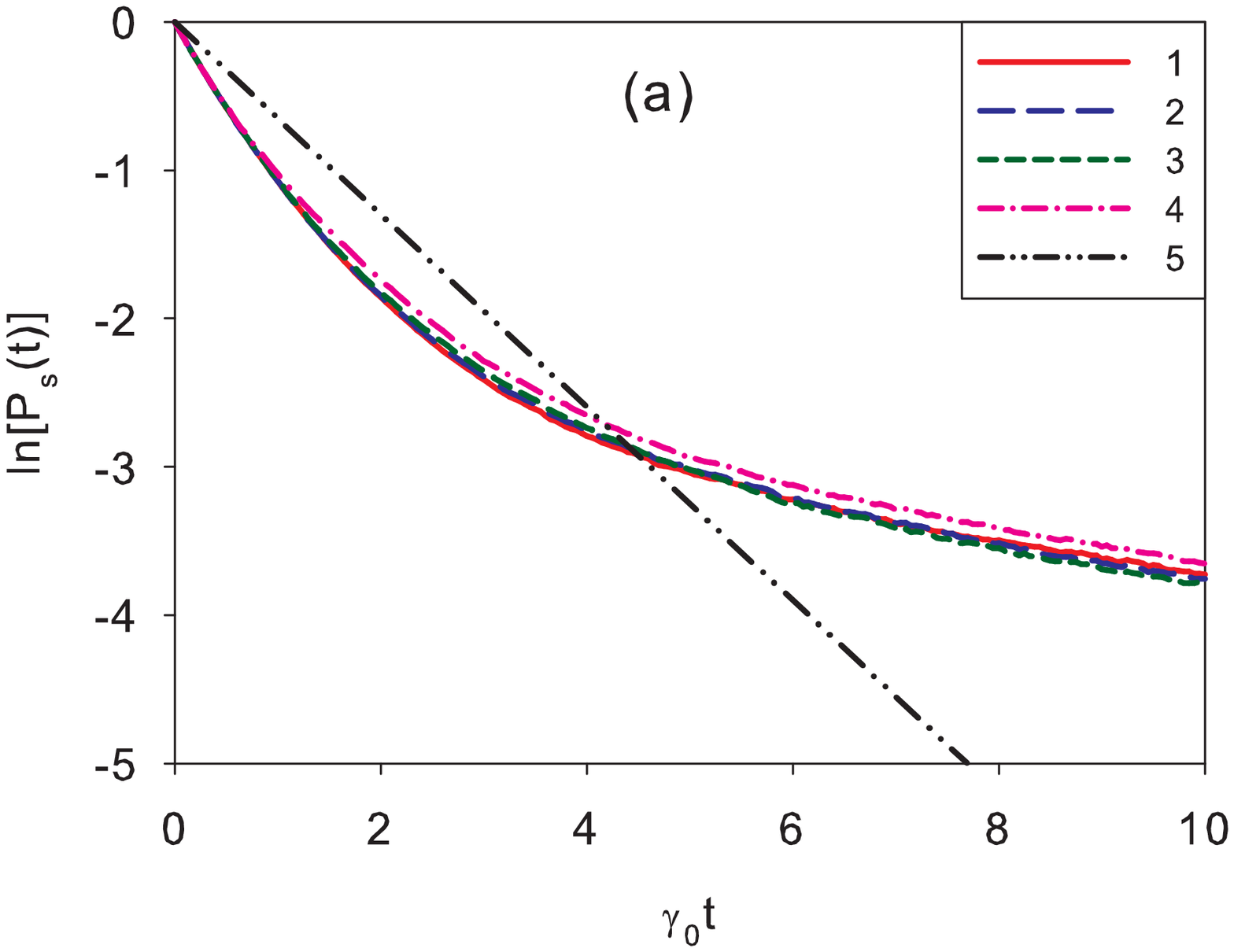}
	\includegraphics[width=6cm]{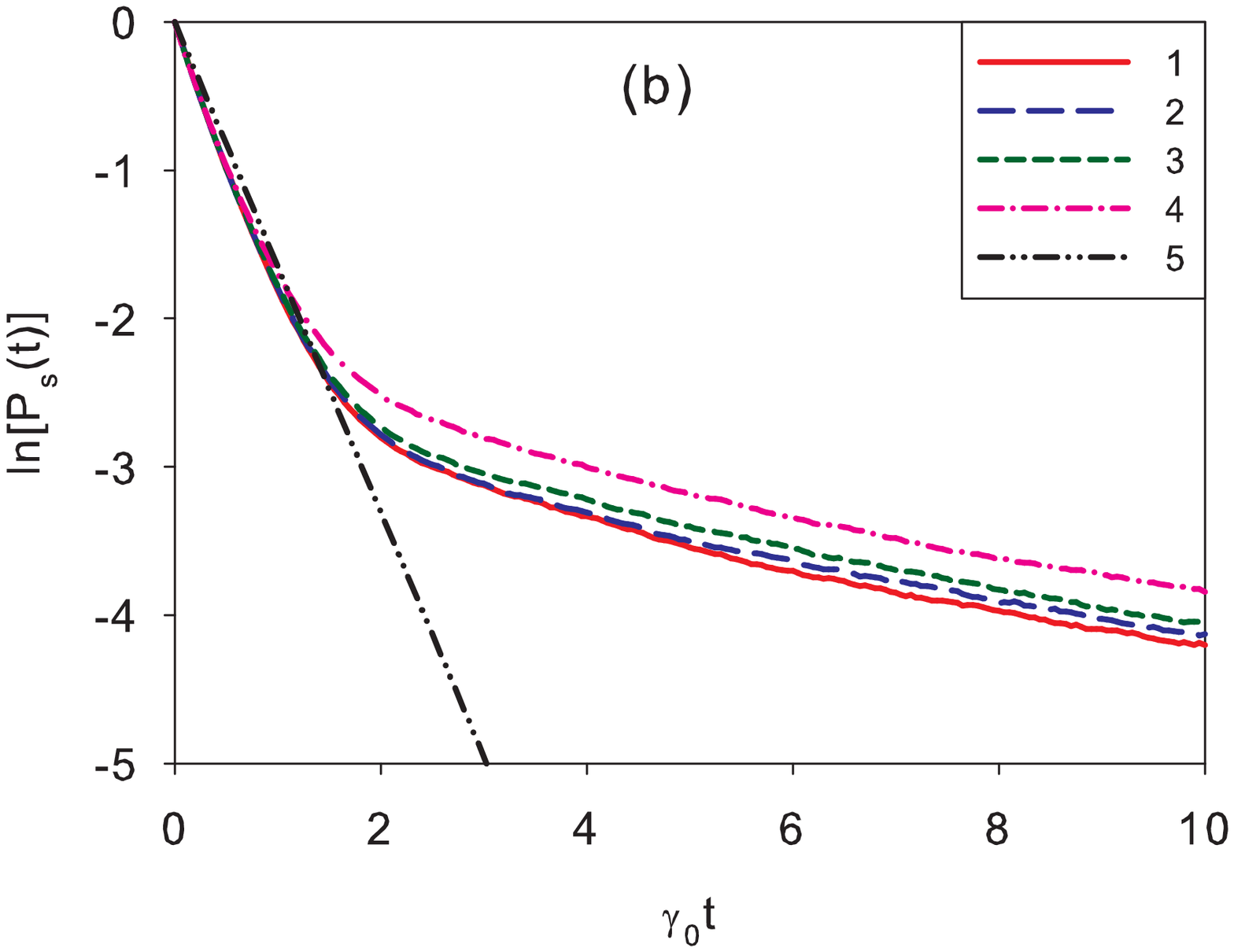}\\
	\caption{\label{fig:two}
			Spontaneous decay dynamics of the initially excited atom. All the parameters are the same as in Fig. \ref{fig:one}. (a) $m=\pm1$; (b) $m=0$. 1, $\Delta=0$; 2, $\Delta=0.5\gamma_{0}$; 3, $\Delta=\gamma_{0}$; 4, $\Delta=3\gamma_{0}$; 5, single atom.}\label{f2}
\end{figure}

The inverse Fourier transform of $b_{e}(\omega)$ allows one to obtain
the time dependence of the amplitudes of the onefold
atomic excited states, $b_{e}(t)$. The time-dependent
probability of excitation of any Zeeman sublevel of any atom in an ensemble can be calculated in a standard way: $P_{e}(t)=|b_{e}(t)|^{2}$. Let us analyze the spontaneous decay dynamics of the initially excited atom. Figure \ref{fig:two} shows the decay of the initially populated Zeeman sublevel of the excited state of this atom, $P_{s}(t)$. For comparison, we added the decay dynamics of a single atom near the conducting surface.

In the figure \ref{fig:two} we can see, that the spontaneous decay dynamics of an excited atom located in an ensemble can not be described by a simple one-exponential law like in the case of a single atom. We observe an essential difference between the results corresponding to an atom in an ensemble and a single atom. It is explained by significant recurrent scattering and associated near-field energy
exchange between the atoms. In such a case the dynamics of spontaneous
decay is described by a multi-exponential law. The initial state, where only one atom is excited,
can be expanded over a set of nonorthogonal
eigenvectors of the Green matrix for any random spatial configuration
of the atomic ensemble. There eigenvectors represent collective quantum states of the polyatomic system. Among these collective
states there are both super- and
subradiant ones \cite{Kaiser1} -- \cite{Kaiser3}. The former influences on the dynamics of spontaneous decay predominantly in early stages of the evolution. At short times, an atom located in
an ensemble decays faster than an isolated atom in the free space. With time the role of subradiant quantum
states increases, that is manifested in a decrease of the decay rate.

Furthermore, we see that the results change with increasing in Stark splitting caused by the
electrostatic field. For $m=0$ the influence of the
electrostatic field manifests itself stronger than for $m=\pm1$. In our opinion, this effect can be explained by the circumstances mentioned above (that the spontaneous decay rate for $m=0$ is larger than for $m=\pm1$) and the fact, that in the case of large $\Delta$ the
transition $J=0$ $\leftrightarrow$ $J=1, m=0$ is frequency-separated from the transitions $J=0$ $\leftrightarrow$ $J=1, m=\pm1$, whereas it is no splitting between $m=1$ and $m=-1$.

\section{Conclusion}
We have studied the spontaneous decay of an excited atom located in a dense and disordered ensemble of pointlike impurity atoms embedded in a transparent dielectric and placed near a charged perfectly conducting surface. Our
approach is based on the solution of the non-steady-state
Schrodinger equation for the wave function of the combined system, which
consists of an ensemble of motionless pointlike scatterers (atoms) and weak
electromagnetic field. On the basis of the
general quantum microscopic theory, we have analyzed the simultaneous influence of the surface and the electric field on the transition spectrum of an excited atom located in an ensemble and its spontaneous decay dynamics.

In our opinion, the approach, that we used in this work, can be further applied for the study of light trapping in atomic ensemble located near a charged conducting surface. The time of radiation trapping is determined by the total population of the excited states $P_{sum}(t)$, which can be calculated as a sum
of $|b_{e}(t)|^{2}$ over all the atoms in the whole ensemble. The use of a constant electric field created by a charged surface will propose an exciting tool to control of light trapping, like it has been proved for a static magnetic field \cite{SkipHav2016}. Another way to manage the optical properties of atomic ensembles is the use of AC control field, which is quasiresonant to one of the atomic transitions, in particular, under conditions of electromagnetically induced transparency \cite{Dat2006} -- \cite{Dat2008}, coherent population trapping \cite{Lit2011} -- \cite{Bar3} or double radio-optical resonance \cite{dror}.

\ack This work was partially supported by Russian
Foundation for Basic Research (Grant No. 18-32-20022) and the Foundation for the Advancement
of Theoretical Physics and Mathematics "BASIS".
The calculation of the spontaneous decay dynamics (part III, Fig. 2)
was supported by the Russian Science Foundation
(Grant No. 17-12-01085).

\end{document}